\documentclass[superscriptaddress,prl,twocolumn,showpacs,preprintnumbers,amsmath,amssymb,aps,times]{revtex4}

\usepackage{graphics}
\usepackage{graphicx}
\usepackage{epsfig}
\usepackage{amssymb}
\usepackage{color}
\usepackage{dcolumn}%
\usepackage{bm}

\newcommand{\PP}{P}

\begin{document}

\title{Fast computation of multi-scale combustion systems}

\author{Eliodoro Chiavazzo}\email{eliodoro.chiavazzo@polito.it}

\author{Filippo Visconti}\email{filippo.visconti@yahoo.it}

\author{Pietro Asinari}\email{pietro.asinari@polito.it}

\affiliation{Department of Energetics, Politecnico di Torino,
Corso Duca degli Abruzzi 24, 10129 Torino, Italy}

\date{\today}

\begin{abstract}
In the present work, we illustrate the process of constructing a simplified model for complex multi-scale combustion systems. To this end, reduced models of homogeneous ideal gas mixtures of methane and air are first obtained by the novel Relaxation Redistribution Method (RRM) and thereafter used for the extraction of all the missing variables in a reactive flow simulation with a global reaction model.
\end{abstract}



\maketitle

\section{Introduction and motivation}
Solution of the full set of equations as required in numerical simulations of reactive flows with detailed chemical kinetics represents a quite challenging task even for today super-computers. The one reason is the large number of kinetic equations needed for tracking each chemical species. On
the other side, detailed combustion mechanisms are typical {\em multi-scale problems} where different chemical processes, characterized by disparate timescales
ranging over several orders of magnitude (from seconds down to nanoseconds) are present.
As a result, modeling detailed combustion fields comes with a tremendous cost where intensive long simulations are needed to resolve the fastest processes, though one is often interested in the slow dynamics. 
Thus simplification methodologies become, other than highly desirable, mandatory in combustion problems where detailed mechanisms for heavy hydrocarbons (with hundreds chemical species) are used in 2- and 3D simulations.

Notice that, there exist often chemical processes that are much faster than the fluid
dynamic phenomena, so if we are only interested in computing the system
behavior on the time-scale of the fluid mechanics, some chemical processes will
be already equilibrated and thus {\em slaved to} the remaining dynamics. 

In fact, modern simplification techniques are based on a systematic decoupling of the fast equilibrating chemical processes from the rest of the dynamics, and are typically implemented by seeking a \textit{low dimensional
manifold} of slow motions in the solution space of the detailed system.

Much effort has been devoted to setting up such automated model reduction procedures. The method of invariant grids (MIG) \cite{bookGK,GKZ04}, the computational singular perturbation (CSP) method \cite{LG91}, the intrinsic low dimensional manifold (ILDM) \cite{MP92}, the invariant constrained equilibrium edge preimage curve method (ICE-PIC) \cite{RenPope06}, and the method of minimal entropy production trajectories (MEPT) \cite{Leb08} are a few popular techniques.

In this work, we introduce an approximated procedure for the fast computation of detailed combustion fields.
To this end we adopt the novel Relaxation Redistribution Method (RRM) \cite{PHD_THESIS_EC} for the construction of a reduced model of the mechanism GriMech 3.0 describing ideal mixtures of air and methane (53 chemical species, 4 elements and 325 reactions) in a closed system under fixed pressure and mixture-averaged enthalpy. The latter description serves, at a later time, for reconstructing all the missing chemical species in a computationally efficient reactive flow simulation performed with a single-step reaction model.

The paper is organized in sections as follows. The notion of slow invariant manifold and the chemical kinetics equations are briefly reviewed in section \ref{SIM.notion} and \ref{sec.equations}, respectively. The relaxation redistribution method is discussed in section \ref{Overview_relaxation}. A reduced model for air and methane is used in a planar counter-flow flame simulation in section \ref{Rfs}, and conclusions are drawn in section \ref{conclusions}.  

\section{Slow invariant manifold (SIM)} \label{SIM.notion}
In this section, we briefly discuss the notions of {\em slow invariant manifold} for a system of autonomous ordinary
differential equations in a domain $\mathcal{U}$ in $R^n$,
\begin{equation}\label{sys}
\dot{y}=f(y).
\end{equation}
For more details, the interested reader is delegated to the dedicated literature \cite{bookGK,GKZ04}. 
A manifold  $\Omega \subset \mathcal{U}$ is {\it invariant} with respect to the system (\ref{sys}) if inclusion $y(t_0) \in \Omega $ implies that
$y(t) \in \Omega $ for all future time $t > t_0$. 

Equivalently, if the tangent space $T_{\emph{\textbf{y}}}$ to $\Omega$ is defined at $y$, invariance requires: $f(y) \in T_{\emph{\textbf{y}}}.$
In order to transform the latter condition into an equation, it proves convenient to introduce projector operators. Let for any subspace $T_{\emph{\textbf{y}}}$ a
projector $\PP$ onto $T_{\emph{\textbf{y}}}$ be defined with image ${\rm im}\PP
=T_{\emph{\textbf{y}}}$. Then the necessary differential condition
can be expressed by:
\begin{equation}\label{invEq}
(1-\PP)f=0,
\end{equation}
where the left-hand side of equation \ref{invEq} is often called {\it defect of invariance} $\Delta$.
It is worth stressing that, although the notion of invariance discussed above is relatively straightforward, {\em slowness} instead is much more delicate. We just notice that the most part of invariant manifolds are not suitable for model reduction (all semi-trajectory are, by a definition, 1D invariant manifold). In this respect, we should also point out that {\em slow manifolds} are not uniquely defined in the literature and, in general, different methods delivers different objects.

Here, we follow the approach of the Method of Invariant Grid (MIG) \cite{bookGK,ChGoKa07}, where slowness is intended as {\em stability}, thus SIM are the
{\it stable stationary solution} of a relaxation process ({\em film equation})
\begin{equation}\label{film} 
\frac{{\rm d} F(\xi)}{{\rm d} t}=(1-\PP) f.
\end{equation} 

We notice that the unknown of both
(\ref{invEq}) and (\ref{film}) is the manifold $\Omega$, which can be conveniently represented in a parametric form, as mapping 
$F:\mathcal{W} \to \mathcal{U}$ of a domain $\mathcal{W}$
(reduced space or parameter space in the following) into the phase-space $\mathcal{U}$; $\Omega$ being the image of this mapping: $\Omega = F(\mathcal{W})$.
\section{Reaction kinetics equations} \label{sec.equations}
\begin{figure}
	\centering
		\includegraphics[width=0.45\textwidth]{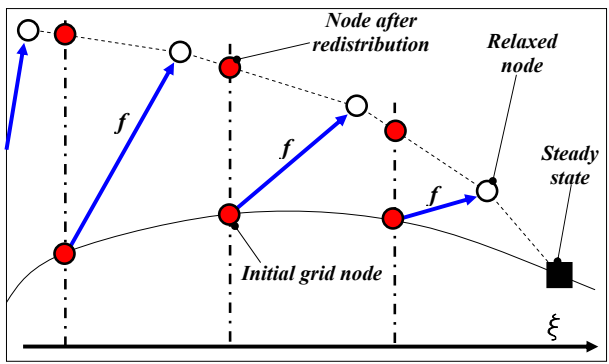}
	\caption{Basic idea behind the Relaxation Redistribution Method.}\label{RRM.principle}
\end{figure}
Here, we consider closed reactive systems with fixed mixture-averaged enthalpy $\bar h$, and total pressure $p$ where $n$ chemical species and $d$ elements participate in a complex network of elementary reactions. 
Species compositions are represented in terms of mass fractions $Y_i=m_i/m_{tot}$, with $m_i$ and $m_{tot}$ denoting the mass of species $i$ and the total mass, respectively. The mixture enthalpy, at a temperature $T$, can be expressed as $\overline h=\sum_{i=1}^{n}{h_{i}(T)Y_{i}}$, while the governing equations in a closed reactor take the form \cite{Turns}:
%
%
%
\begin{equation}\label{ydot}
\dot y=f(y)=(\frac{\dot{\omega}_{1}W_{1}}{\bar \rho},...,\frac{\dot{\omega}_{n}W_{n}}{\bar \rho})^{T},
\end{equation}
where $\bar \rho$ is the mixture density, while $W_i$, $\dot \omega_i$, $h_i(T)$ denote the molecular weight the molar concentration rate and specific enthalpy of species $i$, respectively. Following ChemKin \cite{Chemkin}, the specific enthalpy can be approximated by a polynomial fit as follows:  
%
\begin{equation}\label{poly}
h_{i}(T)=RT(a_{1i}+\frac{a_{2i}T}{2}+\frac{a_{3i}T^{2}}{3}+\frac{a_{4i}T^{3}}{4}+\frac{a_{5i}T^{4}}{5}+\frac{a_{6i}}{T}),
\end{equation}
where $a_{ji}$ are the tabulated Nasa coefficient and $R$ the universal gas constant. Molar concentration rates take the explicit form:
\begin{equation}
\dot \omega_i=\sum_{j=1}^{s}{\left( \beta_{ij}-\alpha_{ij} \right) r_j},
\end{equation}
where $\alpha_{ij}$ and $\beta_{ij}$ are the stoichiometric coefficients of the $j$-th elementary reaction $\sum_{i=1}^{n} \alpha_{ij} X_i \rightleftharpoons \sum_{i=1}^{n} \beta_{ij} X_i$. The $j$-th reaction rate $r_j$ is expressed using the popular {\em mass action law}:
\begin{equation}
r_j = k_j^+ \prod_{i=1}^{n} \left[ X_i \right]^{\alpha_{ij}} - k_j^- \prod_{i=1}^{n} \left[ X_i \right]^{\beta_{ij}}, 
\end{equation}
with $\left[ X_i \right]$ denoting the molar concentration of the $i$th species and $k_j^{\pm}$ the $j$-th reaction rate constant typically expressed in the Arrhenius form \cite{Turns}.
\section{Model reduction technique} \label{Overview_relaxation}
%
%
%
In the following, we use a discrete representation for manifolds referred to as {\it grids} \cite{GKZ04}, consisting of a set of {\em interconnected} nodes, where it is assumed that the nearest neighbors of an arbitrary node $y$ can be identified. 
A grid is defined by the restriction of mapping $F$ on the discrete subset of the parameter space $\mathcal{G}$$\subset \mathcal{W}$ into the phase space $\mathcal{U}$, whereas and {\em invariant grid} satisfies the grid version of the invariance equation: $f(F(\xi))-\PP f(F(\xi))=0, \; \forall \xi\in \mathcal{G}$ \cite{GKZ04}. Notice that, thanks to the node {\em connectivity}, it is possible to compute local tangent space hence the projector $\PP$ (e.g. using approximated differential operators).

\subsection{Relaxation methods}\label{RM}
Here, construction of one-dimensional invariant grids is accomplished by the Relaxation Redistribution Method (RRM) which has proven an efficient method for solving the film equation (\ref{film}) starting from an initial grid $\mathcal{G}_0$. The interested reader can find further details in \cite{PHD_THESIS_EC}.
 Referring to Fig. \ref{RRM.principle}, for simplicity, in this work $\mathcal{G}_0$ is chosen regular in terms of the parameter $\xi$. 

Let a numerical scheme (Euler, Runge-Kutta, etc.) be chosen for solving the system of kinetic equations (\ref{ydot}), and let all the grid nodes relax towards the slow invariant manifold (SIM) under the detailed dynamics $f$ during one time step. The fast component of $f$ brings a grid node closer to the SIM while, at the same time, the slow component causes a contraction towards the steady state of (\ref{ydot}). As a result, the grid becomes dense in a neighborhood of the steady state and coarse far from it, when keeping relaxing. The slow motion can be neutralized by a node redistribution after the grid relaxation (thus mimicking the term $-\PP f$ in (\ref{film})). In other words, as illustrated in Fig. \ref{RRM.principle}, the relaxed states are redistributed on a regular grid in terms of the parameter $\xi$ via linear interpolation. 

Notice that, all intermediate grids are, by construction, regular in terms of $\xi$ and, in the case of an invariant grid, the overall effect due to relaxation and redistribution is null and the invariance condition satisfied. 

In our study, we consider a detailed combustion mechanism for air and methane (GriMech 3.0), where $n=53$ chemical species and 4 elements participate in 325 elementary reactions \cite{GriMech30_short}.
\begin{figure} 
	\centering
		\includegraphics[width=0.45\textwidth]{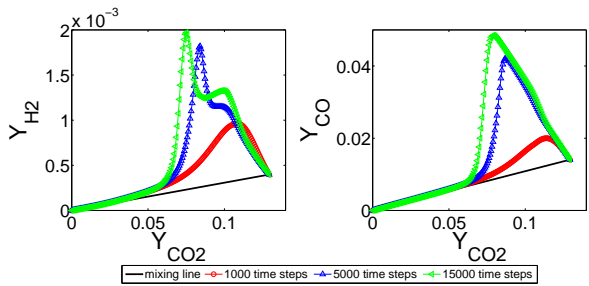}
	  \caption{Relaxation redistribution method where an explicit first order Euler scheme is adopted during relaxation with an adaptive time step $\delta t \le 1.5\times10^{-8}$.}\label{H2}
\end{figure}
Here, at a fixed mixture-averaged enthalpy $\bar h$ and pressure $p$, $\mathcal{G}_0$ represents the {\em mixing line} between the two states $y^{fresh}$ (stoichiometric fresh mixture) and $y^{eq}$ (stoichiometric chemical equilibrium state) discretized by $N=200$ nodes. 
Iterations have carried out until $\delta y / \delta \tilde{y} < \varepsilon$ at every grid node, where $\delta y$ is the overall movement  due to the relaxation and redistribution while $\delta \tilde y$ is the movement due to relaxation alone of an arbitrary node $y$ 
with a tolerance $\varepsilon=0.001$.
\begin{figure} [htbp]
	\centering
		\includegraphics[width=0.48\textwidth]{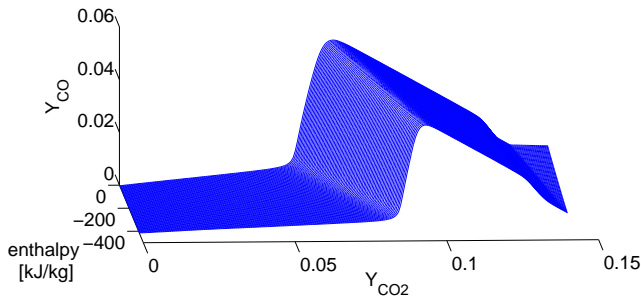}
	\caption{$Y_{CO}$ coordinate. Two-dimensional invariant grid via the relaxation redistribution method.}\label{LUT_CO}
\end{figure}
To the end of constructing a two-dimensional invariant grid parameterized with respect to $\xi=Y_{CO2}$ and $\overline h$, the above construction is performed over a range of enthalpies $-415$ [$kJ/kg$]  $< \bar h <$ $-5$ [$kJ/kg$] with a step $\Delta\overline{h}=-5$ [$kJ/kg$]. In Figure \ref{LUT_CO}, a projection of the above invariant grid in the three-dimensional sub-space $\overline{h}-Y_{CO2}-Y_{CO}$ is reported.



\section{Reactive flow simulation} \label{Rfs}
\begin{figure}[ht]
	\centering
		\includegraphics[width=0.40\textwidth]{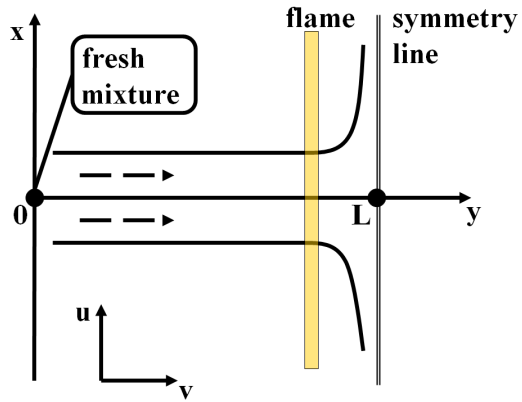}
	\caption{Planar counter-flow configuration. Symmetry is assumed with respect to the line $y=L$.}\label{flame}
\end{figure}
Let us consider the planar stagnation point flow, where a well premixed stoichiometric mixture of air and fuel, initially at room condition ($T=300 K$, $p=1bar$), impinges against a stream of hot products. Due to symmetry, a flat flame can be established in this flow at $0<y<L$ as schematically depicted in Figure \ref{flame}. 
Although the above is effectively a two dimensional problem,
 under the assumptions of symmetry, boundary layer approximation and low Mach number regime, it is possible to consider the following one dimensional system of governing equations imposing conservation of mass, momentum, energy and chemical species, respectively, along the symmetry-line ($x=0$):
\begin{equation}\label{mass_3}
\frac{\partial \bar \rho}{\partial t}+\frac{\partial V}{\partial y}+ \bar \rho U \epsilon = 0,
\end{equation}
\begin{equation}\label{momentum_3}
\bar \rho \frac{\partial U}{\partial t} + \bar \rho U^{2} \epsilon + V \frac{\partial U}{\partial y} - \frac{\partial}{\partial y}(\mu \frac{\partial U}{\partial y}) - \bar \rho^{fresh} \epsilon=0,
\end{equation}
\begin{equation}\label{energy_3}
\bar \rho \frac{\partial T}{\partial t} + V \frac{\partial T}{\partial y} - \frac{1}{\bar C_{p}} \frac{\partial}{\partial y}(\lambda \frac{\partial T}{\partial y}) - \sum_{i=1}^{n}{\frac{\dot{\omega}_{i} h_{i}}{\bar C_{p}}} = 0,
\end{equation}
\begin{equation}\label{species_3}
\bar \rho \frac{\partial Y_{i}}{\partial t} + V \frac{\partial Y_{i}}{\partial y} - \frac{\partial}{\partial y}(\bar \rho D_{i} \frac{\partial Y_{i}}{\partial y})-\dot{\omega}_{i} W_{i} = 0, \quad i=1,...,n,
\end{equation}
where the ideal gas law, $\bar \rho=p \bar W/RT$, can be used for the closure. Moreover, $\mu$, $\lambda$, $D_i$ and $\bar \rho^{fresh}$ are the dynamic viscosity, thermal conductivity, the diffusion coefficient of species $i$ and the density of the fresh mixture at the inlet, respectively. The mean specific heat $\bar C_p$ (under constant pressure) and the mean molecular weight $\bar W$ take the explicit form:
\begin{equation}\nonumber
 \bar C_p = \sum_{i=1}^{n} c_{pi} Y_i, \quad \bar W = \frac{1}{\sum_{i=1}^{n} Y_i/W_i}.
\end{equation}
with $c_{pi}$ being the specific heat of species $i$ (mass unit). 
Let $u$, $v$ and $\epsilon$ be the two velocity components of 2D flow field along the $x$ and $y$ axes and the flame strain rate, respectively. We note that the above set of equations (\ref{mass_3})-(\ref{species_3}) are conveniently expressed in terms of the quantities $U=u/u_{\infty}$ and $V=\bar \rho v$, with $u_{\infty}=\epsilon x$. The latter problem is solved imposing fresh mixture condition at the inlet ($y=0$):
\begin{equation}\label{boundary_y0}
y=0: \quad U=1, \; T=300[K], \; Y_i=Y_i^{fresh},
\end{equation}
while chemical equilibrium and zero-flux condition can be chosen at the outlet 
\begin{equation}\label{boundary_yL}
y=L: \quad U = \sqrt{\frac{\bar \rho^{fresh}}{\bar \rho^{eq}}}, \; V=0, \; \frac{\partial T}{\partial y}=0, \; \frac{\partial Y_i}{\partial y}=0,
\end{equation}
where $\bar \rho^{eq}$ is the density of the fully burned mixture.

The detailed derivation of the set of equations (\ref{mass_3})-(\ref{species_3}) along with the boundary conditions (\ref{boundary_y0}) and (\ref{boundary_yL}) can be found in \cite{MarzoukMAST}.
\begin{figure}[ht]
	\centering
		\includegraphics[width=0.53\textwidth]{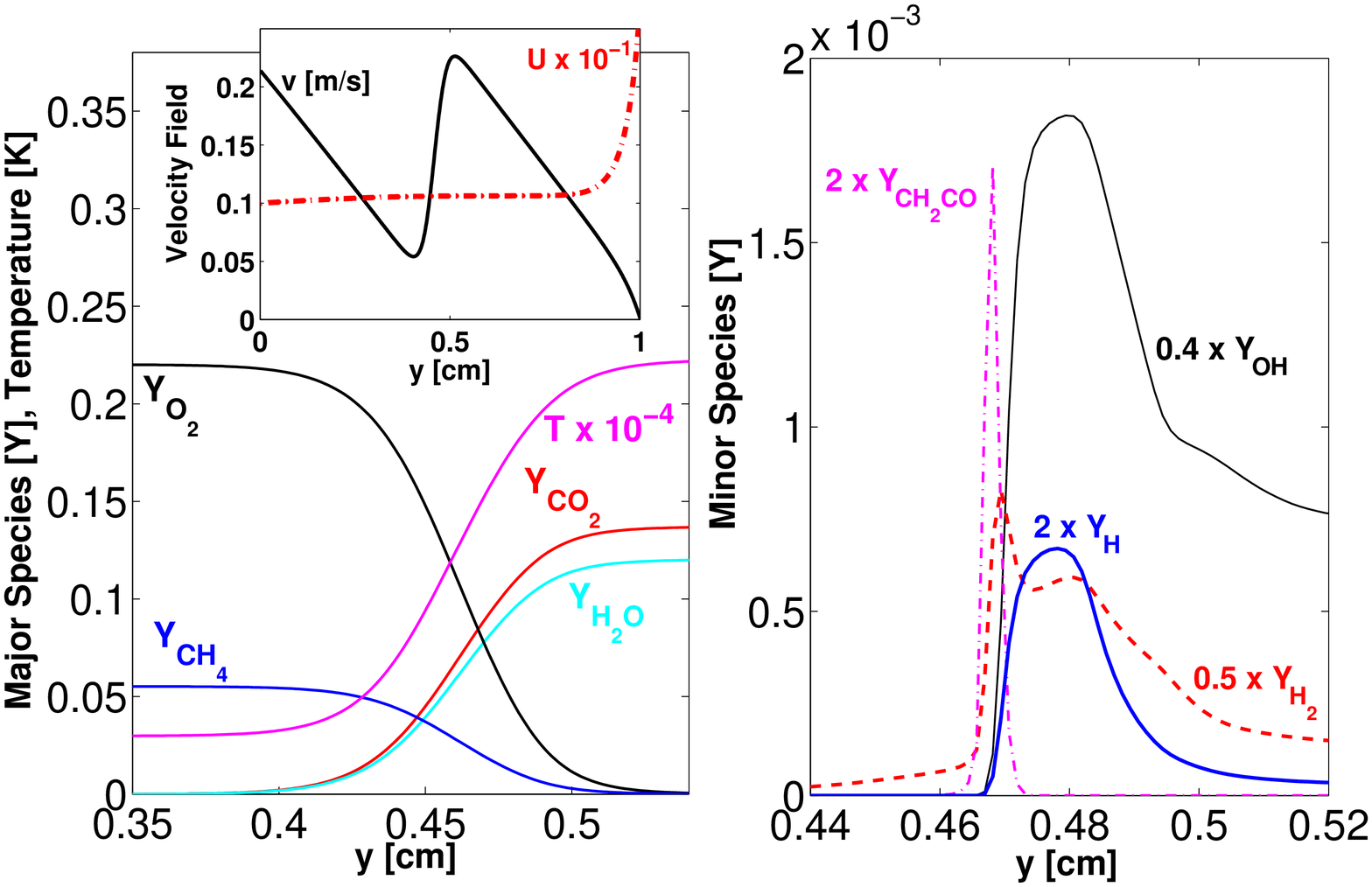}
	\caption{Premixed flame in a planar counter-flow. Equations (\ref{mass_3})-(\ref{species_3}) are solved by finite difference scheme using the global reaction model (\ref{global.reaction}) and (\ref{global.rate}). Minor chemical species are thereafter computed accessing the invariant grid in Figure \ref{LUT_CO}. Here, we use $D_i=D=3 \times 10^{-5}[m^2s^{-1}]$, $\lambda=0.026[Wm^{-1}K^{-1}]$, fixed $p=1[bar]$ and strain rate $\epsilon=40[s^{-1}]$.}\label{complessivo}
\end{figure}
In this study, spatial derivatives in (\ref{mass_3})-(\ref{species_3}) have been approximated by finite differences (upwind for convective terms while central differences for diffusive terms), and the corresponding ordinary differential equations (ODE) system  
has been solved by a numerical stiff solver \cite{ode15s} readily available in Matlab$\textsuperscript{\textregistered}$ ({\em ode15s}). Moreover, we first consider $n=4$ reactive chemical species ($CH_4$, $CO_2$, $O_2$, $H_2O$) with an abundant inert ($N_2$) participating in the one-step global oxidation ($s=1$):  
\begin{equation}\label{global.reaction}
CH_4 + 2 O_2 + 7.52 N_2 \rightarrow CO_2 + 2 H_2O +7.52 N_2,
\end{equation}
whose rate $r_1$, according to \cite{Turns}, can be evaluated as follows:
\begin{equation}\label{global.rate}
r_1=-1.3 \times 10^{8} e^{-24358/T} \left[CH_4\right]^{-0.3} \left[O_2\right]^{1.3}.
\end{equation}
\subsection{Missing fields retrieval}
Upon the solution of the governing equations (\ref{mass_3})-(\ref{species_3}) in combination with the global reaction mechanism (\ref{global.reaction}), five chemical fields, the temperature, the mixture density and flow velocity are available along the symmetry line in Figure \ref{flame}. On the one hand, the computational cost of the latter simulation is drastically reduced compared to a reactive simulation with a detailed combustion mechanism, where a much stiffer and larger set of species equations (\ref{species_3}) must be taken into account. However, on the other hand, global approaches inevitably come with a significant lack of information concerning all intermediate and minor chemical species which underlie a complex phenomenon such as the one represented by (\ref{global.reaction}).    

In order to fill this gap, here we suggest to perform an {\em a posteriori} retrieval of all the missing variables in the above computation (e.g., minor species such as radicals) via linear interpolation from the table describing the invariant grid evaluated (once and forever) in section \ref{Overview_relaxation} by means of the detailed combustion mechanism GriMech 3.0 \cite{GriMech30_short} ($n=53$ and $s=325$). To this end, the two variables $Y_{CO_2}$, $\bar h$ can be extracted from the above simulation and used to access any of the coordinates of the invariant grid (see also Figure \ref{LUT_CO}). Some of the interpolated variables, in a well developed flame along the channel, are reported in Figure \ref{complessivo}.    

\section{Conclusion}\label{conclusions}
In this work, we first demonstrate that a recently introduced model reduction technique (Relaxation Redistribution Method - RRM) is suitable for handling a complex multi-scale combustion mechanism for hydrocarbons. Moreover, on the basis of the latter simplified model, we introduce and test an embarrassingly simple method for the computation of minor chemical species upon a reactive flow simulation efficiently performed with a global (not necessarily one-step) reaction. The present study is the first step towards fast computation of detailed combustion fields for heavy hydrocarbons in 2D and 3D problems.    

\end{document}